\documentclass[reprint,aps,ams,amsmath,prx,twocolumn,longbibliography,superscriptaddress]{revtex4-1}
\usepackage{graphics}
\usepackage{graphicx}
\usepackage{epsfig}
\usepackage{amsmath}
\usepackage{amssymb}
\usepackage{amsfonts}
\usepackage{bm}
\usepackage{color}
\usepackage{bbm}
\usepackage{hyperref}
\usepackage[normalem]{ulem}
\usepackage{comment}
\usepackage{soul}

\newcommand{\beg}{\begin{equation}}
\newcommand{\en}{\end{equation}}
\newcommand{\veps}{\varepsilon}
\newcommand{\eps}{\epsilon}
\newcommand{\bp}{\mathbf p}

\newcommand{\bk}{\mathbf k}
\newcommand{\br}{\mathbf r}

\newcommand{\bn}{\mathbf n}

\newcommand{\dg}{^\dagger}
\newcommand{\St}{\mathrm{St}}

\begin{document}

\title{Resonant second harmonic generation in a two-dimensional electron system}

\author{Maxim Dzero}
\affiliation{Department of Physics, Kent State University, Kent, Ohio 44242, USA}

\author{Jaglul Hasan}
\affiliation{Department of Physics, University of Wisconsin-Madison, Madison, Wisconsin 53706, USA}

\author{Alex Levchenko}
\affiliation{Department of Physics, University of Wisconsin-Madison, Madison, Wisconsin 53706, USA}

\date{April 16, 2025}

\begin{abstract}
We consider the nonlinear response of a disordered two-dimensional electronic system, lacking inversion symmetry, to an external alternating electric field. The application of an in-plane static magnetic field induces local contributions to the current density that are quadratic in the electric field and linear in the magnetic field. This current oscillates at twice the frequency of the external irradiation and there are two linearly independent vector combinations that contribute to the current density.  This particular mechanism coexists with the topological Berry-dipole contribution to the second harmonic of the current density, which can be generated by quantum confinement. Additional nonlocal terms in the current density are possible in the regime away from the normal incidence. The total current exhibits a nonreciprocal character upon reversal of the magnetic field direction. We evaluate the magnitude of this effect by computing its dependence on the strength of spin-orbit coupling and the disorder scattering rate. Importantly, we show that these local second-harmonic contributions can be resonantly excited when the frequency of the external radiation approaches the energy separation between the spin-orbit split bands.
\end{abstract}

\maketitle

\section{Introduction}

The continuing interest in nonlinear optical phenomena is driven by advances in optical measurement techniques \cite{THz1, THz2}, as well as by the peculiar physics arising from the topological properties and nonreciprocity of quantum materials, as reviewed in Refs. \cite{Morimoto:2016, Tokura:2018, Yanase:2024}. In particular, nonlinear effects such as second harmonic generation in normal conductors and third harmonic generation in superconductors have become focal points of theoretical and experimental studies.
For instance, in conventional superconductors, the application of terahertz pulses can excite both amplitude and phase modes \cite{Armitage2023} and induce the inverse Faraday effect -- i.e., the emergence of static magnetization caused by circularly polarized light \cite{IFESC1, IFESC2, IFESC3} (for a recent discussion, see also Ref. \cite{dzero2024ife} and references therein).

In the nonlinear optical response, the electron current density $\mathbf{j}$ can be phenomenologically expanded in powers of the driving electric field $\mathbf{E}$.
In a centrosymmetric material, the leading nonlinear contribution starts at the cubic order, $\mathbf{E}^3$, unless the field is inhomogeneous.
In such cases, quadratic terms are also possible, but they must include spatial gradients. Examples of allowed terms include 
$\mathbf{E}(\mathbf{\nabla}\cdot\mathbf{E})$ and $(\mathbf{E}\cdot\mathbf{\nabla})\mathbf{E}$ \cite{Jha:1965}. Depending on whether the field is static or alternating, various effects can emerge. From the cubic term, one can obtain phenomena such as static nonlinear conductivity, the Kerr effect, photocurrent, dc-induced second harmonic generation, and third harmonic generation \cite{Butcher:1991,Ivchenko:2005}.

\begin{figure}[t!]
\includegraphics[width=0.95\linewidth]{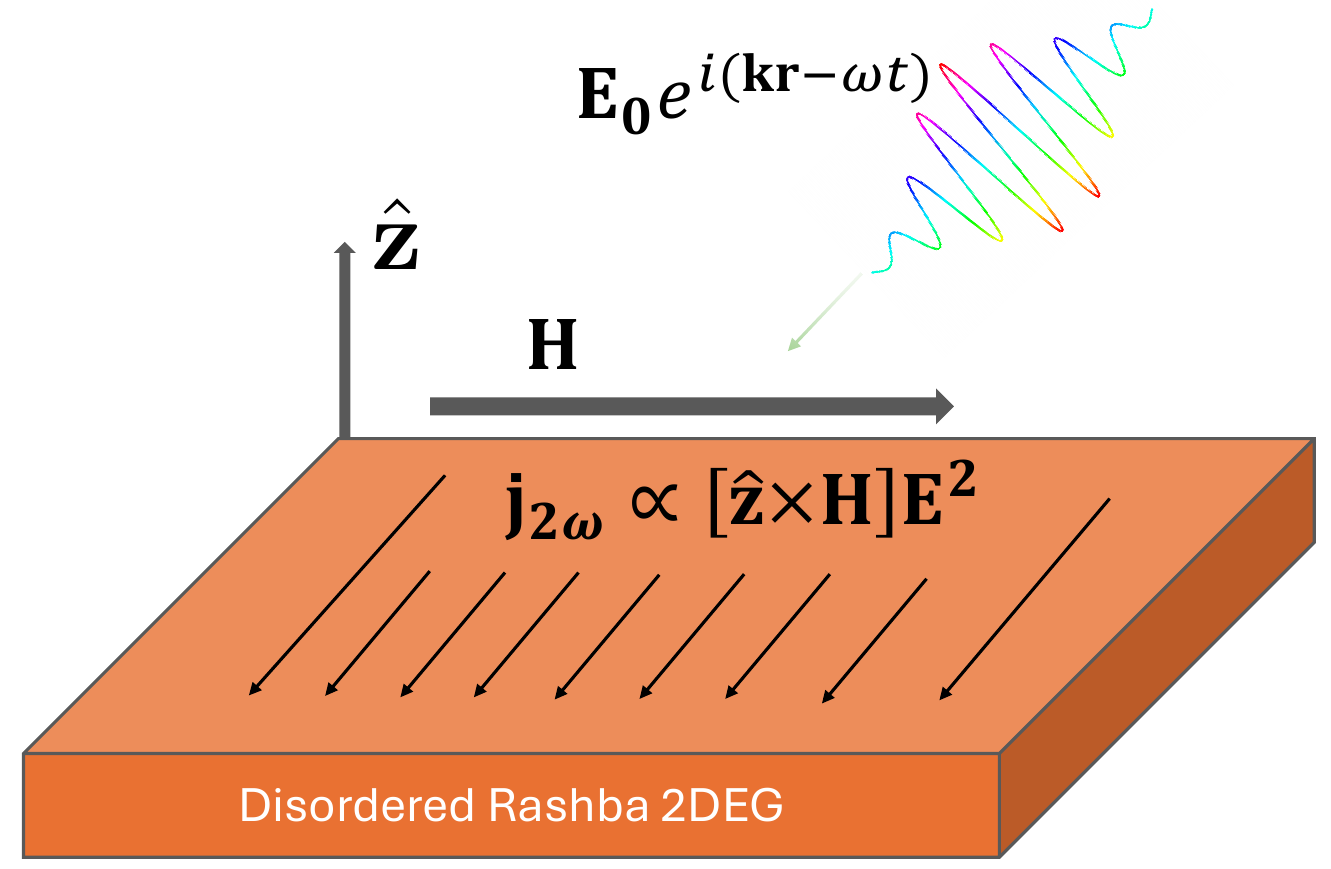}
\caption{Schematic of the system under consideration: electrons are confined to a quantum well, forming a two-dimensional electron gas (2DEG) with Rashba-type spin-orbit interaction. The 2DEG is represented pictorially as a slab. The system is illuminated by light of frequency $\omega$ and subjected to an in-plane static magnetic field $\mathbf{H}$. The combination of these two stimuli induces a second harmonic of the electrical current density, which flows in a direction defined by the vector product of the out-of-plane spin-orbit axis and the magnetic field. This current scales proportionally with the optical power illuminating the 2DEG.}
\label{fig:SHG-2DEG}
\end{figure}

In contrast, in materials whose symmetry group includes a polar axis -- common in those lacking inversion symmetry -- the electron current density may contain terms that are quadratic in the field, even in a homogeneous field. Consequently, the nonlinear part of the electron current density can be characterized by a rank-three tensor:
\begin{equation}
\mathbf{j}_{\text{a}}=\sigma_{\text{abc}}(\omega_1,\omega_2)\mathbf{E}_{\text{b}}(\omega_1)\mathbf{E}_{\text{c}}(\omega_2),
\end{equation}
which represents the second-order nonlinear optical conductivity $\sigma_{\text{abc}}$. These materials exhibit a variety of intriguing physical phenomena, such as the photogalvanic and piezoelectric effects \cite{Sturman1992,Belinicher1980,Belinicher1982,Sipe-Shkrebtii}.

Indeed, for $\omega_1=-\omega_2=\omega$, the product of the fields can be decomposed into symmetrized terms, $\propto \sigma_{\text{abc}}(\mathbf{E}_{\text{b}}\mathbf{E}^*_{\text{c}}+\mathbf{E}^*_{\text{b}}\mathbf{E}_{\text{c}})$, describing linear photogalvanic effects (LPGE), and antisymmetrized terms, $\propto i\varsigma_{\text{ab}}[\mathbf{E}\times\mathbf{E}^*]_{\text{b}}$, describing circular photogalvanic effects (CPGE). The microscopic origin of these effects can be traced to the asymmetry of electronic scattering processes (for a review, see Ref. \cite{Sturman1984} and references therein). For matched frequencies, $\omega_{1}=\omega_2=\omega$ the $\mathbf{E}^2$ term gives rise to second harmonic generation (SHG).

In materials with topologically nontrivial band structures, finite Berry curvature contributes to unique photocurrents \cite{Deyo:2009,Moore:2010,Juan:2017,Konig:2017}, the quantum nonlinear Hall effect \cite{Sodemann:2015,Konig:2021}, and more generally, gyrotropic Hall effects \cite{Konig:2019}. Some of these effects have been observed, such as in semiconductor heterostructures \cite{Ganichev:2001, Olbrich:2009} and in Weyl semimetals like TaAs \cite{Wu:2017,Torchinsky:2018}, which exhibit a giant nonlinear optical response -- orders of magnitude higher than that of archetypal electro-optical materials such as GaAs \cite{Bergfeld:2003} and ZnTe \cite{Wagner:1998}.

In this work, we focus on the specific mechanism of second harmonic generation (SHG) in a two-dimensional electron gas (2DEG). To this end, we consider a 2D electron system described by the Rashba model of spin-orbit coupling \cite{Rashba:1959,Bychkov:1984} and account for electron-impurity scattering, as illustrated in Fig. \ref{fig:SHG-2DEG}. We demonstrate that the application of an in-plane magnetic field induces an oscillatory current response, which exhibits a resonant structure at frequencies commensurate with the energy separation between the spin-orbit split bands. The direction of the induced nonlinear current is determined by a linear superposition of two vectors one of which points along the direction of external electric field while the other points along the direction given by the cross product of magnetic field and normal to the surface. 

\section{Qualitative discussion}

There are several important factors that need to be discussed in relation to the problem at hand. Setting aside model-specific assumptions, it is important to emphasize that in realistic scenarios photocurrents, as it has been shown in Ref. \cite{Moore:2010}, can be generated even in the absence of a magnetic field due to topological effects arising from quantum confinement. In 2D systems, the Berry curvature $\mathbf{\Omega}$ has only one nonvanishing component, which points out of the plane. Consequently, the corresponding Berry curvature dipole $\mathbf{D}$ lies in the 2D plane. This vector defines a specific direction in the plane, giving rise to a finite in-plane current $\mathbf{j}_{2\omega}$ as discussed in Ref. \cite{Sodemann:2015}. At the simplest level, using the Boltzmann equation within the relaxation-time approximation, the second harmonic of the current density can be expressed as: 
\begin{equation}\label{eq:SHG-Berry-dipole} 
\mathbf{j}_{2\omega}=\frac{e^3\tau(\mathbf{D}\cdot\mathbf{E})}{2(1+i\omega\tau)}[\hat{\mathbf{z}}\times\mathbf{E}], 
\end{equation} where $\hat{\mathbf{z}}$ is the unit vector along the normal to the plane of motion, 
$\tau$ is the elastic scattering time due to point-like impurities (we adopt the units $\hbar=1$ throughout the paper). As a result, this current as a function on frequency exhibits a Lorentzian shape. It is worth noting that other extrinsic contributions from disorder, such as skew-scattering and side-jump effects \cite{Deyo:2009, Moore:2010,Konig:2017}, also contribute to $\mathbf{j}_{2\omega}$, though their role in this effect has not been thoroughly explored. Note that we are not presenting here additional topological interband contributions to the SHG as characterized by the shift vector $\mathbf{R}$, which is defined with Berry connections \cite{Baltz:1981,Morimoto:2016}. 

The effect of spin-orbit interaction alone would not yield a finite $\mathbf{j}_{2\omega}$, as the nonlinear response should still vanish in the local limit. However, as argued in Ref. \cite{Edel1988}, breaking time-reversal symmetry with an in-plane magnetic field, which selects a definite direction within the layer, generates a finite current. In the clean limit, and at sufficiently high frequencies exceeding the spin-splitting energy, it was found that 
\begin{equation}\label{eq:SHG-spin-orbit}
\mathbf{j}_{2\omega}=-\frac{2m}{\pi}\left(\frac{e\alpha_{\textrm{so}}}{\omega}\right)^3\frac{g\mu_{\text{B}}}{\omega}[\hat{\mathbf{z}}\times{\mathbf H}]{\mathbf E}^2, 
\end{equation}   
where $m$ is the electron effective mass, $g$ is the gyromagnetic factor, $\alpha_{\textrm{so}}$ is the Rashba spin-orbit coupling constant and $\mu_{\text{B}}$ is the Bohr magneton. The fact that the current density can contain only odd powers of the spin-orbit coupling constant arises from general symmetry considerations under inversion transformation. The comparison of this contribution to Eq. \eqref{eq:SHG-Berry-dipole} in the regime $\omega\tau\gg1$ corresponding to a clean limit, makes it clear that Eq. \eqref{eq:SHG-spin-orbit} displays a much faster decay in frequency. 
   
In what follows, we revisit the calculation of $\mathbf{j}_{2\omega}$ presented in Ref. \cite{Edel1988} and generalize the results obtained earlier for a clean metallic layer to the case of disordered system assuming the limit of weak electron-electron interactions.  For this purpose, we develop a consistent microscopic approach to analyze the corresponding physical effects by deriving the quantum kinetic equation for the Wigner distribution function (WDF). The equation for the WDF is constructed to describe the evolution of the electron system in the presence of disorder in two spatial dimensions under an external monochromatic radiation \cite{MSH2004,Andrey2006}. 

We compute the local correction to the second harmonic of the current density, determining its dependence on the spin-orbit coupling strength, disorder scattering rate, and the frequency of external radiation. Several vector combinations contribute to $\mathbf{j}_{2\omega}$. For instance, one term proportional to $\propto\mathbf{E}(\mathbf{E}\cdot[\hat{\mathbf{z}}\times\mathbf{H}])$ and another $\propto(\mathbf{E}\cdot\mathbf{H})[\hat{\mathbf{z}}\times\mathbf{E}]$ persist even in the clean limit, however there are only two linearly independent terms that determine the magnitude and direction of the total current density.

Perhaps the most significant aspect of the problem, which has not been addressed before, concerns the role of inter-band transitions in electron scattering. We find that all contributions to the second harmonic current density can be resonantly excited when the radiation frequency matches the energy splitting between the two Rashba-split bands. For example, near the resonance, $|\omega-\omega_{\text{res}}|\ll\omega_{\text{res}}$ with $\omega_{\text{res}}=2\Delta$, we find instead of Eq. \eqref{eq:SHG-spin-orbit} the following asymptotic result 
\begin{equation}\label{eq:SHG-res}
\mathbf{j}_{2\omega}\approx\frac{3m}{8\pi}\frac{(e\alpha_{\textrm{so}})^3}{(\omega_{\text{res}}-\omega-i\gamma)^3}\left(\frac{g\mu_{\text{B}}}{\Delta}\right)[\hat{\mathbf{z}}\times{\mathbf H}]{\mathbf E}^2, 
\end{equation}   
where $\Delta$ is the characteristic energy scale of spin-orbit coupling, and $\gamma=\tau^{-1}$ is the energy width of the disorder-induced broadening. 
This resonance enhances the feasibility of observing the SHG experimentally. Additionally, Eq. (\ref{eq:SHG-res}) implies that the total current, including the topological terms, will display nonreciprocity -- meaning that the magnitude of the induced current differs when the direction of the magnetic field is reversed. On the other hand, the kinematic processes involving intra-band scattering produce second-harmonic contributions of the order 
$O(g\mu_{\text{B}}H/\veps_{\text{F}})$, where $\veps_{\text{F}}$ is the Fermi energy. For this reason, these contributions can be safely neglected due to their small magnitude.


\section{Formulation of the model and technical approach}

\subsection{Hamiltonian}

We begin with the following model Hamiltonian, which describes noninteracting electrons constrained to move in a plane and subjected to an external alternating vector potential $\mathbf{A}$ and a static in-plane magnetic field $\mathbf{H}$:
\beg\label{Eq1}
\begin{split}
&\hat{\cal H}(\br,t)=\frac{1}{2m}\left[\hat{\bp}-\frac{e}{c}{\mathbf A}(\br,t)\right]^2\nonumber\\
&+\alpha_{\textrm{so}}\left(\hat{\mathbf{z}}\times\hat{\mbox{\boldmath $\sigma$}}\right)\cdot\left(\hat{\bp}-\frac{e}{c}{\mathbf A}(\br,t)\right)
+g\mu_{\text{B}}({\mathbf H}\cdot\hat{\mbox{\boldmath $\sigma$}})+U(\br).
\end{split}
\en
Here, $\hat{\bp}=-i{\mbox{\boldmath $\nabla$}}$ is the electron momentum, $\hat{\mbox{\boldmath $\sigma$}}$ is the set of Pauli matrices, $U(\br)$ is the disorder potential, and ${\mathbf A}(\br,t)$ is the vector potential corresponding to the periodically modulated electric field ${\mathbf E}=-(1/c)\partial_t{\mathbf A}={\mathbf E}_0e^{i(\bk\br-\omega t)}+\mathrm{c.c}$. For simplicity, we assume a short-ranged disorder potential generated by point-like impurities whose correlation function is described by a Gaussian white noise 
\beg\label{CorrDis}
\left\langle U(\br)U(\br')\right\rangle=\frac{\delta(\br-\br')}{2\pi\nu_{\text{F}}\tau},
\en
where angular brackets denote disorder averaging. The disorder scattering rate is $\tau^{-1}$ and $\nu_{\text{F}}=m/2\pi$ is the single-particle density of states (per spin). 

\subsection{Keldysh form of the Dyson equation}

We are interested in computing the nonequilibrium current density of the system governed by the model Hamiltonian Eq. \eqref{Eq1}. For this purpose we introduce the retarded $\hat{G}^R$, advanced $\hat{G}^A$, and Keldysh $\hat{G}^K$ Green's functions which are $2\times 2$ matrices in the spin space and they satisfy the Dyson equations \cite{Kamenev2009,Kamenev2011}
\beg\label{Eq2}
\left(\hat{\cal G}_0^{-1}-\check{\Sigma}\right)\circ\check{G}=\check{1}, \quad \check{G}\circ\left(\hat{\cal G}_0^{-1}-\check{\Sigma}\right)=\check{1}, 
\en
where the Green's function matrix is  
\beg
\quad \check{G}=\left(\begin{matrix} \hat{G}^R & \hat{G}^K \\ 0 & \hat{G}^A \end{matrix}\right),
\en
and $\hat{\cal G}_0^{-1}=i\partial_{t}-\hat{\cal H}(\br,t)$. The self-energy part is governed by disorder scattering and is determined within the constraints of the self-consistent Born approximation (SCBA)
\beg
\check{\Sigma}(x,x')=\frac{\delta(\br-\br')}{2\pi\nu_{\text{F}}\tau}\check{G}(x,x'),
\en
where we used the shorthand notation $x=(\br,t)$. 

\begin{figure}[t!]
\includegraphics[width=0.95\linewidth]{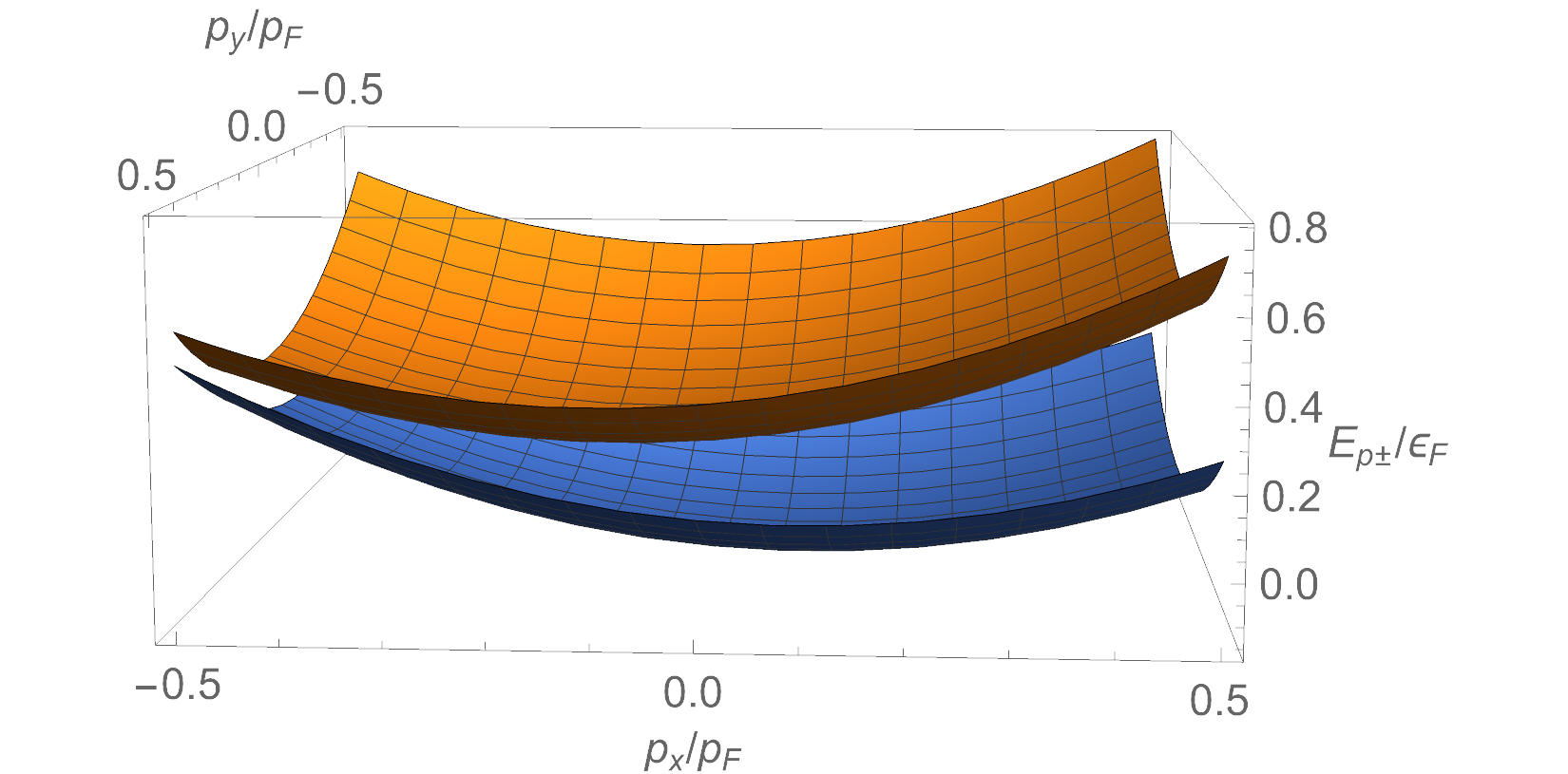}
\caption{Energy bands of the electron spectrum in the presence of spin-orbit coupling and an in-plane magnetic field, plotted according to Eq. \eqref{NewEnergyBands}. To generate this plot, we normalized energies by the Fermi energy and momenta by the Fermi momentum. We used the following values for the dimensionless parameters: $\alpha_{\text{so}}/v_{\text{F}}=0.1$ and $h/\varepsilon_{\text{F}}=0.15$. In the plot, the angle between the direction of $\mathbf{h}$ and $p_x$ is set to $\pi/6$.}
\label{fig:E-Bands}
\end{figure}

In the absence of electron-electron interactions and external alternating electromagnetic fields, the retarded and advanced components of $\check{G}(x,x')$ depend only on $x-x'$. Given the form of the Hamiltonian Eq. \eqref{Eq1}, after performing a Fourier transform on these functions, one readily finds:
\beg\label{GRGA}
\hat{G}_{\bp\eps}^{R(A)}=\frac{\hat{\Pi}_\bp^{(+)}}{\eps-E_{\bp,+}\pm \frac{i}{2\tau}}+\frac{\hat{\Pi}_\bp^{(-)}}{\eps-E_{\bp,-}\pm\frac{i}{2\tau}},
\en
where 
\beg\label{NewEnergyBands}
E_{\bp,\pm}=\frac{p^2}{2m}\pm\sqrt{\alpha_{\textrm{so}}^2p^2+h^2+2\alpha_{\textrm{so}}([\hat{\mathbf{z}}\times\mathbf{h}]\cdot\bp)}
\en
are the energies of the two branches of the electron spectrum, and  $\hat{\Pi}_\bp^{(\pm)}=(1\pm{{\mathbf N}_\bp}\cdot{\mbox{\boldmath $\hat{\sigma}$}})/2$, are the projection operators onto these branches, with the notations ${\mathbf N}_\bp={\mathbf b}_\bp/b_\bp$, and ${\mathbf b}_\bp=\alpha_{\textrm{so}}[\bp\times\hat{\mathbf{z}}]+\mathbf{h}$
(we also introduced $\mathbf{h}=g\mu_{\text{B}}{\mathbf H}$ and $h=|\mathbf{h}|$, as well as $b_\bp=|\mathbf{b}_\bp|$ for brevity). In Fig. \ref{fig:E-Bands}, we plot the energy bands for a specific choice of parameters, as an illustration.

The information about the nonequilibrium state of a system is contained in the function $\hat{G}^K$. Using the Dyson equation (\ref{Eq2}) it is straightforward to show that the Keldysh component of $\check{G}(x,x')$ satisfies the relation 
$\hat{G}^K=\hat{G}^R\circ\hat{\Sigma}^K\circ\hat{G}^A$. From this relation we derive the following equation for the Keldysh function:
\beg\label{Eq4GK}
\left[\hat{G}^R\right]^{-1}\circ\hat{G}^K-\hat{G}^K\circ\left[\hat{G}^A\right]^{-1}=\hat{\Sigma}^K\circ\hat{G}^A-\hat{G}^R\circ\hat{\Sigma}^K.
\en
At this point, it is customary to proceed by performing the Wigner transformation of Eq. \eqref{Eq4GK} with respect to the `center-of-mass' coordinates, $X=(x+x')/2$, and the relative coordinates, $\delta x=x-x'$:
\beg\label{WignerDefine}
\hat{G}^K(x,x')=\int\frac{d\veps d^2\bp}{(2\pi)^3}\hat{g}^K_{\bp\veps}(X)e^{i(\bp+(e/c){\mathbf A}(X))(\br-\br')-i\veps(t-t')}.
\en
to derive the corresponding kinetic equation for the semiclassical Green's function $\hat{g}^K_{\bp\veps}$. This can be done by borrowing the methods develped in the context of the spin Hall effect, see e.g. Refs. \cite{MSH2004, Dzero2022}. However, we find that while this approach is valid, it is not particularly suited for computing the nonlinear response to an alternating electric field due to the explicit time dependence of the retarded and advanced Green's functions. Instead, in this paper, we will use an alternative but similar approach, based on the analysis of the quantum kinetic equation for the Wigner distribution function \cite{Andrey2006}. We believe this approach is more advantageous, as it allows for the consideration of various types of disorder potentials and is more tractable for calculating transport quantities such as current density.

\subsection{Wigner distribution function}

To this end, we introduce the Wigner distribution function (WDF) for the electrons which is a $2\times 2$ matrix in spin space and is defined according to 
\begin{align}\label{WDF}
W_{\alpha\beta}(\bk,\eps;\br,t)=\frac{1}{2\pi}\int d^2{\mathbf s}\int d\tau e^{i\bk{\mathbf s}-i\eps\tau}\nonumber \\ 
\left\langle\psi_\beta\dg\left(\br+\frac{\mathbf s}{2},t+\frac{\tau}{2}\right)\psi_\alpha\left(\br-\frac{\mathbf s}{2},t-\frac{\tau}{2}\right)\right\rangle.
\end{align}
In order to derive the kinetic equation for the WDF, it is useful to recall the definition of the Green's functions on the Keldysh contour
\beg\label{Gij}
i\hat{G}_{\alpha\beta}^{(kl)}(1,1')=\langle \hat{T}_t\psi_\alpha(1_k)\psi_\beta\dg(1_l')\rangle, \quad (k,l=1,2).
\en
Here $1$ is a short notation for the spatial and temporal coordinates, $l=1$ refers to the upper branch of the Keldysh contour which runs from $-\infty$ to $+\infty$, while $l=2$ refers to the lower branch of the Keldysh contour which runs from $+\infty$ to $-\infty$. From this definition it follows that WDF can be expressed in terms of the `lesser' correlation function
\beg\label{WDFG12}
W_{\alpha\beta}(\bk,\br;\eps,t)=-\frac{i}{2\pi}\hat{G}_{\alpha\beta}^{(12)}(\bk,\br;\eps,t).
\en
Given the relation 
\beg\label{G12}
\hat{G}_{\bk\eps}^{(12)}(\br,t)=\frac{1}{2}\left(\hat{G}_{\bk\eps}^{K}(\br,t)-\hat{G}_{\bk\eps}^{R}(\br,t)+\hat{G}_{\bk\eps}^{A}(\br,t)\right),
\en
one can employ Eq. \eqref{WDFG12} along with the expressions in Eq. \eqref{GRGA} for the retarded and advanced functions to obtain WDF in equilibrium 
\beg\label{weqso}
W_{\alpha\beta}^{(0)}(\bk,\eps)=\frac{1}{2}\sum\limits_{s=\pm}\left(\delta_{\alpha\beta}+s\frac{{\mathbf b}_\bk\cdot{\mbox{\boldmath $\sigma$}}}{b_\bk}\right)
f(\eps)\delta\left(\eps-E_{\bk s}\right),
\en
where $f(\eps)$ is the Fermi-Dirac distribution function. More generally, one can write down an equation for the WDF using equations for the retarded and advanced functions, which follow from the Dyson equations (\ref{Eq2}) and equation for the Keldysh function \eqref{Eq4GK}. Since we are primarily interested in the local response, we will ignore the dependence of the WDF on $\br$. Collecting all the relevant contributions we obtain the following equation:
\begin{align}\label{Eq4w}
&\left(\partial_t+\frac{1}{\tau}\right)\hat{W}_{\bk\eps}(t)+i\left\{({\mathbf b}_\bk\cdot{\mbox{\boldmath $\hat{\sigma}$}});\hat{W}_{\bk \eps}(t)\right\}_{-}=\nonumber \\ 
&-\frac{1}{2}\left\{\frac{\bk}{m}+\alpha_{\textrm{so}}\hat{\mbox{\boldmath $\eta$}};\bm{\partial}\hat{W}_{\bk \eps}(t)\right\}_++\St[\hat{W}_{\bk\eps}],
\end{align}
where $\{A;B\}_{\pm}=AB\pm BA$, $\hat{\mbox{\boldmath $\eta$}}=[\hat{\mathbf{z}}\times\hat{\mbox{\boldmath $\sigma$}}]$ and we introduced the shorthand notation
\beg\label{TildeGrad}
\bm{\partial}\hat{W}_{\bk\eps}(t)=\frac{e}{\omega}\left[\hat{W}_{\bk\eps+\frac{\omega}{2}}(t)-\hat{W}_{\bk\eps-\frac{\omega}{2}}(t)\right]{\mathbf E}(t).
\en
The collision integral due to disorder scattering takes the form 
\begin{equation}\label{Icoll}
\begin{split}
&\St[\hat{W}_{\bk\eps}]=\frac{i}{2\pi\nu_{\text{F}}\tau}\\&\times\int\frac{d^2\bk}{(2\pi)^2}
\left[\hat{G}_{\bk\eps}^R(t)\circ\hat{W}_{\bk\eps}(t)-\hat{W}_{\bk\eps}(t)\circ\hat{G}_{\bk\eps}^A(t)\right],
\end{split}
\en
with the convolution that should be understood as follows:
\beg\label{circ}
A_{\bk\veps}(t)\circ B_{\bk\veps}(t)=A_{\bk\veps}(t)e^{\frac{i}{2}(\stackrel{\leftarrow}\partial_\veps\stackrel{\rightarrow}\partial_t-\stackrel{\leftarrow}\partial_t\stackrel{\rightarrow}\partial_\veps)}B_{\bk\veps}(t).
\en
It should be noted that Eq. \eqref{Icoll} is a direct consequence of SCBA. 
Provided that solution for the WDF is found, the charge current density can be computed from  
\beg\label{current}
{\mathbf j}(t)={e}\int\limits_0^\infty\frac{kdk}{2\pi}\int\limits_0^{2\pi}\frac{d\theta}{2\pi}\int\limits_{-\infty}^{\infty}{\mathbf v}_{\bk}(\theta)\textrm{Tr}[\hat{W}_{\bk\eps}(t)]d\eps.
\en 
Here ${\mathbf v}_{\bk}(\theta)=v_k(\cos\theta,\sin\theta)\equiv v_k\bn$ is the bare velocity and $v_k=\partial \eps_k/\partial k$ (see Ref. \cite{Andrey2006} for further details). 

We proceed to solve Eq. \eqref{Eq4w} using perturbation theory. For this purpose, we assume that the disorder scattering rate is sufficiently small. Specifically, we impose the condition $\tau^{-1}\ll\textrm{min}\{\omega,\alpha_{\textrm{so}}k_{\text{F}}\}$, where $k_{\text{F}}$ is the Fermi momentum. This condition allows us to neglect the collision operator (\ref{Icoll}) to leading order. Additionally, we assume that the external magnetic field is sufficiently weak, $g\mu_{\text{B}}H\ll \alpha_{\textrm{so}}k_{\text{F}}$, since we are only interested in the $\mathbf{H}$-linear term in the current density.


\section{Electron current density}

\subsection{Linear response and first harmonic}

We approach the solution of Eq. \eqref{Eq4w} using perturbation theory by successive iterations. Transforming to Fourier space and expanding to first order in the electric field, the solution to the WDF can be compactly expressed as:
 \begin{align}\label{Formalw1}
\hat{W}_{\bk\eps}^{(1)}&=\frac{(z_\omega^2+2b_\bk^2)}{z_\omega(z_\omega^2+4b_\bk^2)}\hat{\cal K}_1+\frac{2b_\bk^2}{z_\omega(z_\omega^2+4b_\bk^2)}\hat{\eta}_\bk\hat{\cal K}_1\hat{\eta}_\bk\nonumber \\ 
&-\frac{ib_\bk}{z_\omega^2+4b_\bk^2}
\left\{\hat{\eta}_\bk;\hat{\cal K}_1\right\}_{-},
\end{align}
where we used notations $\hat{\eta}_\bk=({\mathbf b}_\bk\cdot{\mbox{\boldmath $\hat{\sigma}$}})/b_\bk$, $z_\omega=-i\omega+{1}/{\tau}$, and we also introduced the matrix function 
\beg\label{K1}
\begin{split}
\hat{\cal K}_1&=-\frac{e}{2\omega}\\&\times\left
\{\frac{\bk\cdot{\mathbf E}}{m}+\alpha_{\textrm{so}}({\mathbf E}\times\hat{\mathbf{z}})\cdot{\mbox{\boldmath $\hat{\sigma}$}};
\hat{W}_{\bk \eps+\frac{\omega}{2}}^{(0)}-\hat{W}_{\bk \eps-\frac{\omega}{2}}^{(0)}\right\}_+.
\end{split}
\en
Performing the trace over the spin indices followed by the integration over the directions of $\bk=k\bn$ gives
\begin{equation}\label{Inttheta}
\begin{split}
&\int\limits_0^{2\pi}\frac{d\theta}{2\pi}\bn\textrm{Tr}_\sigma[\hat{W}_{\bk\eps}^{(1)}]=-\frac{e{\mathbf E}\tau}{2\omega (1-i\omega\tau)} \\ 
&\times\sum\limits_{s=\pm}\left(\frac{k}{m}+s\alpha_{\textrm{so}}\right)\left(F_{k\eps+\frac{\omega}{2}}^{(s)}-F_{k\eps-\frac{\omega}{2}}^{(s)}\right),
\end{split}
\end{equation} 
where we used the shorthand notation $F_{k\eps}^{(s)}=f(\eps)\delta(\eps-\eps_k-s\alpha_{\textrm{so}}k)$. It should be noted that the magnetic field dependent corrections to Eq. \eqref{Inttheta} are of the order of $O({\mathbf H}^2)$ and therefore can be neglected. The remaining integration over $\eps$ (cf. Eq. (\ref{current})) becomes trivial and we are left with the following expression for the first harmonic of the current density:
\begin{align}\label{j1w}
{\mathbf j}_{\omega}=-\frac{e^2{\mathbf E}\tau}{2m\omega(1-i\omega\tau)}\sum\limits_{s=\pm}\int\limits_{0}^\infty\frac{k^2dk}{2\pi}
\left(\frac{k}{m}+s\alpha_{\textrm{so}}\right)\nonumber\\ 
\times\left[f\left(\eps_k+s\alpha_{\textrm{so}}k+\frac{\omega}{2}\right)-f\left(\eps_k+s\alpha_{\textrm{so}}k-\frac{\omega}{2}\right)\right].
\end{align}
At zero temperature, $T\to0$, the remaining $k$-integral can be found in elementary functions with the result 
\begin{equation}
{\mathbf j}_{\omega}=\frac{e^2{\mathbf E}\tau}{4\pi m\omega(1-i\omega\tau)}\sum\limits_{s=\pm}\left[\frac{k^4_+-k^4_-}{4m}+\frac{s\alpha_{\text{so}}}{3}(k^3_+-k^3_-)\right],
\end{equation}
where $k_\pm=\sqrt{k^2_{\text{F}}+\alpha^2_{\text{so}}m^2\pm m\omega}-s\alpha_{\text{so}}m$. When the spin-orbit term is neglected, $\alpha_{\text{so}}\to0$, this expression reduces to the standard Drude conductivity at finite frequency,
\begin{equation}
{\mathbf j}_{\omega}=\frac{\sigma_{\text{D}}{\mathbf E}}{(1-i\omega\tau)},\quad \sigma_{\text{D}}=\frac{ne^2\tau}{m}.
\end{equation} 
Here $n=2\nu_{\text{F}}\varepsilon_{\text{F}}$ being the total electron density. 

\subsection{Nonlinear response and second harmonic}

The second order correction to the Wigner distribution function is given by 
\begin{align}\label{Formalw2}
\hat{W}_{\bk\eps}^{(2)}&=\frac{(z_{2\omega}^2+2b_\bk^2)}{z_{2\omega}(z_{2\omega}^2+4b_\bk^2)}\hat{\cal K}_2+\frac{2b_\bk^2}{z_{2\omega}(z_{2\omega}^2+4b_\bk^2)}\hat{\eta}_\bk\hat{\cal K}_2\hat{\eta}_\bk\nonumber \\ 
&-\frac{ib_\bk}{z_{2\omega}^2+4b_\bk^2}\left\{\hat{\eta}_\bk;\hat{\cal K}_2\right\}_{-}, 
\end{align}
where now the second order in electric field perturbation matrix function is determined from
\beg\label{K2}
\begin{split}
&\hat{\cal K}_2=-\frac{e}{2\omega}\\&\times\left\{\frac{\bk\cdot{\mathbf E}}{m}+\alpha_{\textrm{so}}[{\mathbf E}\times\hat{\mathbf{z}}]\cdot{\mbox{\boldmath $\hat{\sigma}$}};
\hat{W}_{\bk \eps+\frac{\omega}{2}}^{(1)}-\hat{W}_{\bk \eps-\frac{\omega}{2}}^{(1)}\right\}_+
\end{split}
\en
and $z_{2\omega}=-2i\omega+{\tau}^{-1}$. In full analogy with the discussion above, we proceed by calculating the trace over the spin indices of $\hat{W}_{\bk\eps}^{(2)}$. For clarity, we separate the expression for $\textrm{Tr}_\sigma[\hat{W}_{\bk\eps}^{(2)}]$ into four parts, each representing a distinct contribution to the second harmonic of the current density. The specific details of this decomposition into different contributions can be found in the Appendices.

The first contribution that we will consider is given by the following expression
\begin{align}\label{Trw2a}
\textrm{Tr}_\sigma\left[\hat{W}_{\bk\eps}^{(2,\textrm{a})}\right]=-\frac{e^2\alpha_{\textrm{so}}^2{\mathbf E}^2}{\omega^2z_\omega z_{2\omega}}\left(\frac{z_\omega^2+2b_\bk^2}
{z_\omega^2+4b_\bk^2}\right)\nonumber \\ 
\times \sum\limits_{s=\pm}\left(2F_{\bk\eps}^{(s)}-F_{\bk\eps+{\omega}}^{(s)}-F_{\bk\eps-{\omega}}^{(s)}\right).
\end{align}
It is evident that the contribution of this term to ${\mathbf j}_{2\omega}$ must be proportional to $[\hat{\mathbf{z}}\times{\mathbf H}]$, as the only dependence on $\bn$ in this expression appears through the combination $\bn\cdot[\hat{\mathbf{z}}\times{\mathbf H}]$, which is part of the definition of $b_\bk$. Expanding Eq. (\ref{Trw2a}) to linear order in the magnetic field and performing the remaining integrations over $\theta$, $\eps$ and $k$, we obtain (see Appendix \ref{app:A} for details):
\beg\label{j2a}
{\mathbf j}_{2\omega}^{(\textrm{a})}=-A(\omega)\frac{2m}{\pi}\left(\frac{e\alpha_{\textrm{so}}}{\omega}\right)^3\frac{g\mu_{\text{B}}}{\omega}[\hat{\mathbf{z}}\times{\mathbf H}]{\mathbf E}^2,
\en
where the dimensionless function $A$ is defined as follows 
\beg\label{Awtau}
A(\omega)=2(\omega\tau)^4\frac{(1-i\omega\tau)}{(1-2i\omega\tau)}
\frac{(1-i\omega\tau)^2-4\zeta^2}{[(1-i\omega\tau)^2+4\zeta^2]^3}
\en
and $\zeta=\alpha_{\textrm{so}}k_{\text{F}}\tau$ is the emergent dimensionless parameter of the model. Structurally, Eq. \eqref{j2a} takes the same form as the result previously obtained for the second harmonic of the current density in a clean metal, provided that we set $A(\omega)\to1$ (see Eq. \eqref{eq:SHG-spin-orbit} in Ref. \cite{Edel1988}). However, in the clean limit, $\tau\to\infty$, this function approaches $A(\omega)\approx\omega^4(\omega^2+4\Delta^2)/(\omega^2-4\Delta^2)^3$, where $\Delta=\alpha_{\textrm{so}}k_{\text{F}}$ represents the characteristic energy scale of spin-orbit coupling. Thus, $A(\omega)$ assymptotes to a unity only at sufficiently high frequencies, $A(\omega\gg\Delta)\approx 1$. A key feature missing in this high-frequency limit, assumed in Ref. \cite{Edel1988}, is the resonance at $\omega=2\Delta$. This resonance suggests that kinematic processes contributing to $\hat{W}_{\bk\eps}^{(2,\textrm{a})}$ must necessarily involve electron transfer between spin-orbit split bands. At finite scattering time $\tau$, the resonant frequency shifts by an amount determined by $\zeta$, while the resonance width broadens, proportional to $\tau^{-1}$.

We proceed with the second contribution, which is given by the following expression
\begin{align}\label{Trw2b}
\!\!\!\textrm{Tr}_\sigma\!\left[\hat{W}_{\bk\eps}^{(2,\textrm{b})}\right]&\!=\!-\frac{e^2(\bk\cdot{\mathbf E})}{m\omega^2z_\omega z_{2\omega}}\!\!
\sum_{s=\pm}\!\!\left(2F_{\bk\eps}^{(s)}-F_{\bk\eps+{\omega}}^{(s)}-F_{\bk\eps-{\omega}}^{(s)}\right)\nonumber \\ 
&\times\left(\frac{\bk\cdot{\mathbf E}}{m}+\frac{2s\alpha_{\textrm{so}}}{b_\bk}[{\mathbf E}\times\hat{\mathbf{z}}]\cdot{\mathbf b}_\bk\right).
\end{align}
In order to extract a nonvanishing contribution to the second harmonic of the current density, we expand $b_\bk$ to first order in the magnetic field. 
Performing the necessary integrations yields an expression for the second harmonic contribution ${\mathbf j}_{2\omega}^{(\textrm{b})}$, which is parametrically smaller than ${\mathbf j}_{2\omega}^{(\textrm{a})}$ as it is proportional to $g\mu_{\text{B}}H/\veps_F$ (see Appendix \ref{app:B} for details). For this reason, this contribution can be neglected. It is noteworthy that the processes defining ${\mathbf j}_{2\omega}^{(\text{b})}$, unlike those contributing to ${\mathbf j}_{2\omega}^{(\textrm{a})}$, do not involve the resonant enhancement, i.e. the contributions to $\hat{W}_{\bk\eps}^{(2,\textrm{b})}$ are governed by the kinematic processes which involve electron scattering processes within each spin-orbit split band. 

Two additional terms contribute to $\textrm{Tr}_\sigma[\hat{W}_{\bk\eps}^{(2)}]$, and due to their similarities, we discuss them together as they 
only differ by a sign and a vector structure of the pre-factor.  We find from Eq. \eqref{Formalw2} 
\begin{align}\label{Trw2c}
\textrm{Tr}_\sigma\left[\hat{W}_{\bk\eps}^{(2,\textrm{c})}\right]=-\frac{2e^2\alpha_{\textrm{so}}^2([{\mathbf E}\times\hat{\mathbf{z}}]\cdot{\mathbf b}_\bk)^2}{\omega^2z_\omega z_{2\omega}(z_\omega^2+4b_\bk^2)}
\nonumber \\ 
\times\sum\limits_{s=\pm}\left(2F_{\bk\eps}^{(s)}-F_{\bk\eps+{\omega}}^{(s)}-F_{\bk\eps-{\omega}}^{(s)}\right),
\end{align}
and also 
\begin{align}\label{Trw2d}
\textrm{Tr}_\sigma\left[\hat{W}_{\bk\eps}^{(2,\textrm{d})}\right]=\frac{2e^2\alpha_{\textrm{so}}^2([{\mathbf E}\times\hat{\mathbf{z}}]\times{\mathbf b}_\bk)^2}{\omega^2z_\omega z_{2\omega}(z_\omega^2+4b_\bk^2)}
\nonumber \\ 
\times\sum\limits_{s=\pm}\left(2F_{\bk\eps}^{(s)}-F_{\bk\eps+{\omega}}^{(s)}-F_{\bk\eps-{\omega}}^{(s)}\right).
\end{align}
The field dependence of these traces arises both from the explicit dependence on ${\mathbf b}_\bk$, which is linear in $\mathbf{H}$, and from the field dependence of the distribution functions. Notably, as in the calculation of ${\mathbf j}_{2\omega}^{(\textrm{a})}$, we can disregard the magnetic field dependence in the arguments of the distribution functions in both expressions \eqref{Trw2c} and \eqref{Trw2d}. A detailed analysis in Appendix \ref{app:C} shows that expanding the distribution functions in the magnetic field introduces terms that are of order $O(\alpha_{\text{so}}^2/v_{\text{F}}^2)$ smaller than the leading contributions. With these simplifications, we find that
  \begin{align}\label{j2wc}
{\mathbf j}_{2\omega}^{\textrm{(c)}}&=-\frac{2m}{\pi}\left(\frac{e\alpha_{\textrm{so}}}{\omega}\right)^3\frac{g\mu_{\text{B}}}{\omega}
\left(C(\omega)[\hat{\mathbf{z}}\times{\mathbf H}]{\mathbf E}^2 \right. \nonumber \\ 
&+\left(2C(\omega)-C'(\omega)\right)\left([\hat{\mathbf{z}}\times{\mathbf H}]\cdot{\mathbf E}\right)\mathbf E,
\end{align}
where we have used angular averages 
\begin{align}
&\int \frac{d \hat{\mathbf{n}}}{2 \pi} \hat{\mathbf{n}} (\hat{\mathbf{n}} \cdot \mathbf{E})([\hat{\mathbf{z}} \times \mathbf{H}] \cdot \mathbf{E})=\frac{1}{2} ([\hat{\mathbf{z}} \times \mathbf{H}] \cdot \mathbf{E})\mathbf{E}, \nonumber \\
&\int \frac{d \hat{\mathbf{n}}}{2 \pi} \hat{\mathbf{n}} (\hat{\mathbf{n}} \cdot \mathbf{E})^2([\hat{\mathbf{z}} \times \mathbf{H}] \cdot \hat{\mathbf{n}})=\nonumber \\ 
&\frac{1}{8} \left([\hat{\mathbf{z}} \times \mathbf{H}] \mathbf{E}^2+2([\hat{\mathbf{z}} \times \mathbf{H}] \cdot \mathbf{E})\mathbf{E}\right).
\end{align}
We notice the appearance of the additional vector combinations as compared to ${\mathbf j}_{2\omega}^{\textrm{(a)}}$. The frequency dependent dimensionless functions appearing here are defined as follows:
\begin{subequations}\label{BCwtau}
\begin{align}
C(\omega)&=\frac{4\zeta^2(\omega\tau)^4(1-i\omega\tau)}{(1-2i\omega\tau)[(1-i\omega\tau)^2+4\zeta^2]^3},\\
C'(\omega)&=\frac{2(\omega\tau)^4(1-i\omega\tau)}{(1-2i\omega\tau)[(1-i\omega\tau)^2+4\zeta^2]^2}.
\end{align}
\end{subequations}

From the form of $\textrm{Tr}_\sigma\left[\hat{W}_{\bk\eps}^{(2,\textrm{d})}\right]$, it is evident that the expression for ${\mathbf j}_{2\omega}^{\textrm{(d)}}$ will closely resemble Eq. \eqref{j2wc}. Noting that the vector ${\mathbf b}_\bk$  lies in the $xy$-plane, we have $\left[[{\mathbf E}\times\hat{\mathbf{z}}]\times{\mathbf b}_\bk\right]^2=({\mathbf E}\cdot{\mathbf b}_\bk)^2$. This implies that to obtain the expression for ${\mathbf j}_{2\omega}^{\textrm{(d)}}$, we can formally replace $[{\mathbf E}\times\hat{\mathbf{z}}]$ with ${\mathbf E}$ in (\ref{j2wc}) and change the overall sign. This yields:
\begin{align}\label{j2wd}
{\mathbf j}_{2\omega}^{\textrm{(d)}}&=\frac{2m}{\pi}\left(\frac{e\alpha_{\textrm{so}}}{\omega}\right)^3\frac{g\mu_{\text{B}}}{\omega}
\left(C(\omega) [\hat{\mathbf{z}}\times{\mathbf H}){\mathbf E}^2\right. \nonumber \\ 
&+\left(2C(\omega)-C'(\omega)\right)\left({\mathbf E}\cdot{\mathbf H}\right)
[\hat{\mathbf{z}}\times{\mathbf E}].
\end{align}

\begin{figure}[t!]
\includegraphics[width=\linewidth]{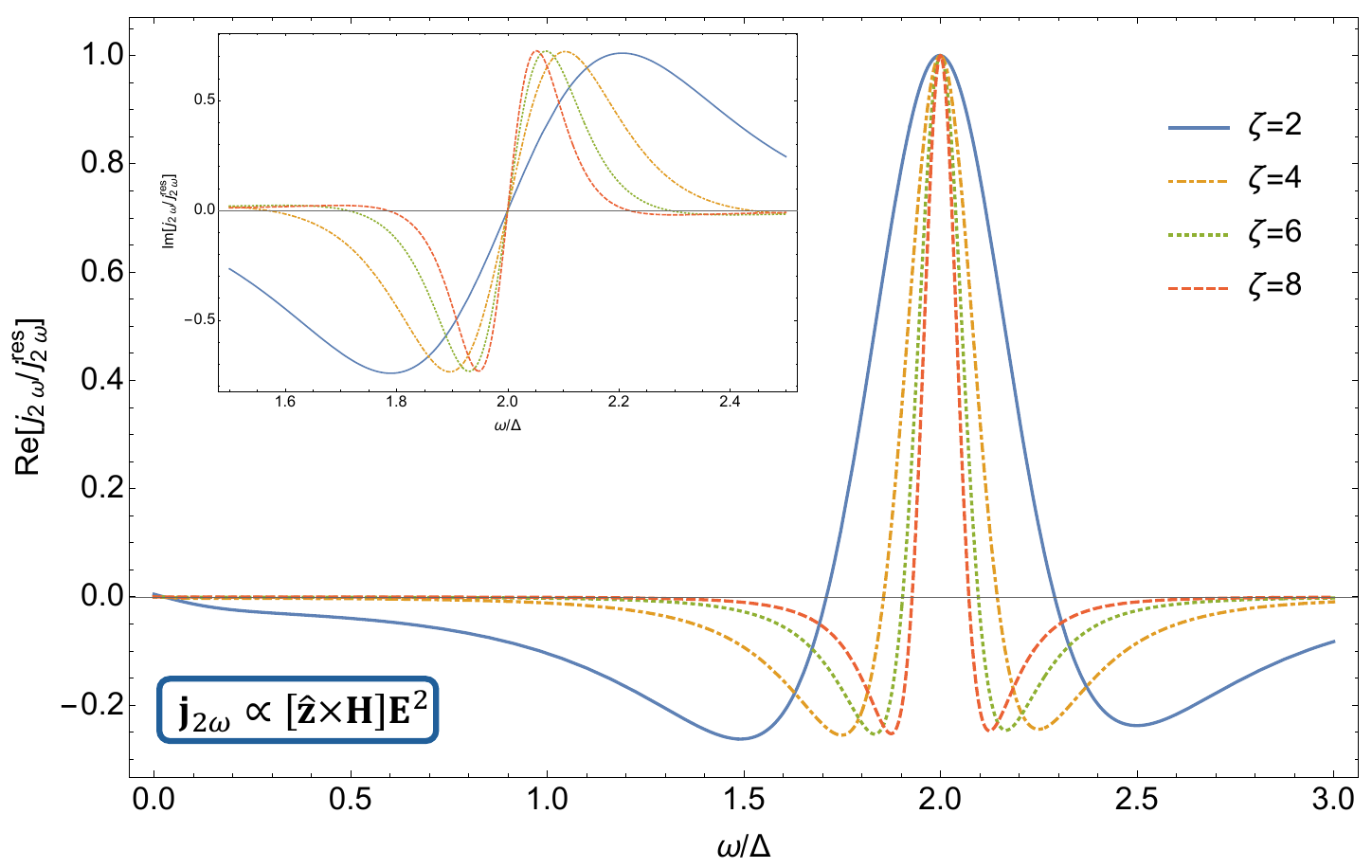}
\includegraphics[width=\linewidth]{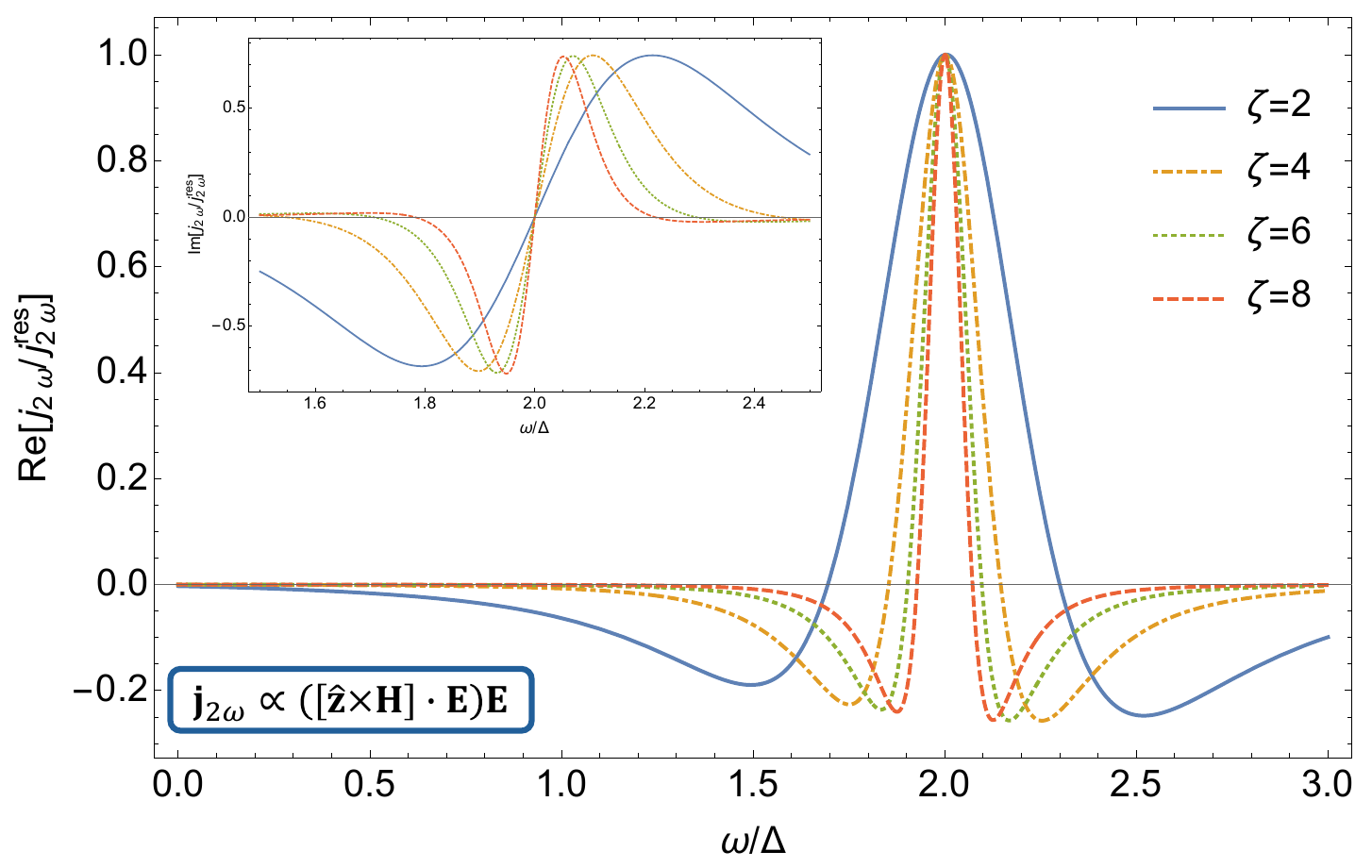}
\caption{Normalized frequency dependence of the second harmonic of the current density, plotted from Eq. \eqref{j2wtotalFIN} in dimensionless units $\omega/\Delta$ and 
 $\zeta=\alpha_{\text{so}}k_{\text{F}}\tau$. Each panel corresponds to a particular vector combination of the current density as marked on the plot.  
The insets show the imaginary part of the same function. The $\zeta$ values shown in the legend are selected to correspond to the ballistic limit, $\zeta>1$.}
\label{fig:Resonance}
\end{figure}

For the total current density, we obtain a particularly compact expression by noticing the following vector identity
\begin{equation}
(\mathbf{E}\cdot[\hat{\mathbf{z}}\times\mathbf{H}])\mathbf{E}+(\mathbf{E}\cdot\mathbf{H})[\hat{\mathbf{z}}\times\mathbf{E}]=[\hat{\mathbf{z}}\times\mathbf{H}]\mathbf{E}^2,
\end{equation}  
which holds for the field orientations in our setup [Fig. \ref{fig:SHG-2DEG}]. The resulting expression can be grouped as 
\begin{align}\label{j2wtotalFIN}
{\mathbf j}_{2\omega}&=-\frac{2m}{\pi}\left(\frac{e\alpha_{\textrm{so}}}{\omega}\right)^3\frac{g\mu_{\text{B}}}{\omega}\nonumber\\
&\times\left\{S(\omega)(\hat{\mathbf{z}}\times{\mathbf H}){\mathbf E}^2+Q(\omega)([\hat{\mathbf{z}} \times \mathbf{H}] \cdot \mathbf{E})\mathbf{E}\right\},
\end{align}
where $S(\omega)=A(\omega)-2C(\omega)+C'(\omega)$ and $Q(\omega)=2(2C(\omega)-C'(\omega))$. This expression represents the main result of our paper.

In Fig. \ref{fig:Resonance} we plot the frequency dependence of the second harmonic with frequency normalized to units of $\Delta=\alpha_{\text{so}}k_{\text{F}}$. 
The resonance is clearly visible at $\omega/\Delta=2$, and the plot shows this for various values of the parameter $\zeta=\tau\Delta$ in the ballistic limit where $\zeta>1$.
Due to the chosen normalization on the pot, the peak height remains fixed. However, in physical units, the maximum at resonance scales as
${\mathbf j}^{\text{res}}_{2\omega}\propto \tau^3$. By expanding $S(\omega)$ near the resonant frequency, using $|\omega-2\Delta|/2\Delta\ll1$ as a control parameter, 
we recover Eq. \eqref{eq:SHG-res} with a broadening $\gamma=\tau^{-1}$. 

\section{Summary and estimates}

In this paper, we have analyzed the nonlinear response of a weakly disordered two-dimensional electronic system exposed to monochromatic electromagnetic radiation in the presence of a static in-plane magnetic field. Using the Keldysh method for out-of-equilibrium systems, we computed the second harmonic contribution to the current density. We find that the effect of second harmonic generation is most pronounced when the frequency of the external electromagnetic wave is tuned to the resonant frequency corresponding to electronic transitions between the two spin-orbit-split bands. Additional contributions to the second harmonic include topological Berry dipole effects and nonlocal effects when the angle of incidence is not exactly normal to the surface. It is useful to assess the relative importance of these mechanisms.

The simplest estimate for the topological term can be derived for a 2D system with $C_s$ symmetry. In this configuration, there is a single component of the Berry curvature aligned along $\hat{\mathbf{z}}$. To lowest order, this expands as $\mathbf{\Omega}=\beta k_x\hat{\mathbf{z}}$, where $\beta$ is a constant with units of volume \footnote{In fact, for a Berry flux linear in $\mathbf{k}$ any linear combination of $k_x$ and $k_y$ can be brought to the above form by a rotation.}. In this scenario, calculating the second harmonic generation using the Boltzmann equation within the relaxation time approximation, following the procedure outlined in Refs. \cite{Deyo:2009,Moore:2010,Sodemann:2015}, yields an estimate for the magnitude of the current density (up to numerical prefactors corresponding to Eq. \eqref{eq:SHG-Berry-dipole}):
\begin{equation}
|\mathbf{j}^{\text{top}}_{2\omega}|\simeq \frac{\beta ne^3\tau}{\sqrt{1+(\omega\tau)^2}}\mathbf{E}^2.
\end{equation}
 The computational tight-binding approach yields a Berry-phase factor $\beta$ on the order of $\beta\sim 1\AA^3$ for various quantum well potentials, with $\beta$ decreasing as the well width increases.
On the other hand, the resonant term corresponding to Eq. \eqref{eq:SHG-res} is
\begin{equation}
|\mathbf{j}^{\text{res}}_{2\omega}|\simeq \frac{m(e\alpha_{\textrm{so}})^3\tau^3}{(1+\delta\omega^2\tau^2)^{3/2}}\left(\frac{g\mu_{\text{B}}H}{\Delta}\right){\mathbf E}^2, 
\end{equation}   
with the notation $\delta\omega=\omega-\omega_{\text{res}}$. Comparing the two at the resonance, $\omega=2\Delta$, noticing that $\tau\Delta\gg1$ for sufficiently clean systems, and requiring $|\mathbf{j}^{\text{top}}_{2\omega}|\lesssim|\mathbf{j}^{\text{res}}_{2\omega}|$, leads to the condition $\beta\lesssim\lambda^3_{\text{F}}(\alpha_{\text{so}}/v_{\text{F}})^3(k_{\text{F}}l)^3(g\mu_{\text{B}}H/\veps_{\text{F}})$, where $l_{\text{F}}=v_{\text{F}}\tau$ is the electron mean free path and $\lambda_{\text{F}}$ is the electron de Broglie wave length. Provided that the smallness in $\alpha_{\text{so}}/v_{\text{F}}$ ratio is compensated by the large dimensionless conductance $k_{\text{F}}l$ so that the product of these factors is approximately of the order of unity, then for the density in the range $n\sim10^{12}$ cm$^{-2}$, corresponding to $\lambda_{F}\sim10^{-8}$ m, the above condition can be securely met even for a relatively small values of $g\mu_{\text{B}}H/\veps_{\text{F}}$. However, due to a strong suppression of the field-induced current at high frequencies, with $\mathbf{j}_{2\omega}\propto 1/\omega^4$, we expect the topological term to dominate at frequencies sufficiently far from the resonance.

An additional competing effect arises if the incidence of the electromagnetic wave on the layer is not perfectly normal. Then the nonlocal contribution to the current should be included, which takes the form 
\begin{equation}\label{nonlocal}
|\mathbf{j}^{\text{nl}}_{2\omega}|\simeq\frac{e^3n}{2m^2\omega^3}|\mathbf{q}_\parallel|{\mathbf E}^2,
\end{equation}
where $\mathbf{q}_\parallel$ is the component of the wave vector $\mathbf{q}$ along the plane of 2DEG. Our estimates show that the non-local current (\ref{nonlocal}) significantly exceeds the local current for nearly all incidence angles at non-resonant frequencies. 

\section*{Acknowledgments}

We express our gratitude to Anton Andreev and Andrey Shytov, whose illuminating discussions were instrumental in shaping our understanding of the physics presented in this paper, and for bringing Ref. \cite{Andrey2006} to our attention. 

This work was financially supported by the National Science Foundation (NSF) grant No. DMR-2002795 (M. D.). J. H. was supported in part by MacDonald award provided by the Department of Physics at the UW-Madison and NSF Grant No. 
DMR-2203411. A. L. acknowledge financial support by the NSF Grant No. DMR- 2452658 and H. I. Romnes Faculty Fellowship provided by the University of Wisconsin-Madison Office of the Vice Chancellor for Research and Graduate Education with funding from the Wisconsin Alumni Research Foundation. The main part of this work has been carried out during the Aspen Center for Physics 2024 Summer Program on ``Probing Collective Excitations in Quantum Matter by Transport and Spectroscopy", which was supported by the National Science Foundation Grant No. PHY-2210452. 

\appendix 

\section{Details of the calculation for ${\mathbf j}_{2\omega}^{(\textrm{a})}$}\label{app:A}

Consider the contribution from the second order correction Eq. (\ref{Trw2a}). We expand the trace $\textrm{Tr}_\sigma[\hat{W}_{\bk\eps}^{(2,a)}]$ up to the linear order in magnetic field using:
\begin{subequations}\label{ExpandFraction}
\begin{align}
&\frac{z_\omega^2+2b_\bk^2}{z_\omega^2+4b_\bk^2}\approx\frac{z_\omega^2+2\Delta_k^2}{z_\omega^2+4\Delta_k^2}-\frac{4g\mu_{\text{B}}\Delta_kz_\omega^2}{(z_\omega^2+4\Delta_k^2)^2}([\hat{\mathbf z}\times{\mathbf H}]\cdot\bn), \\
&f(E_{\bk,s})\approx \left.f(\eps)+sg\mu_{\text{B}}[(\hat{\mathbf z}\times{\mathbf H})\cdot\bk]\left(\frac{df_\eps}{d\eps}\right)\right|_{\eps=\eps_k+s\Delta_k}.
\end{align}
\end{subequations}
Here we introduced the notation $\Delta_k=\alpha_{\textrm{so}}k$ and $s=\pm$. In order to compute the current, we need to integrate over the directions of $\bk$. Angular averages can be found by using the standard integrals  
\begin{align}\label{angular}
\int\limits_0^{2\pi}\frac{d\theta}{2\pi}n_in_j&=\frac{1}{2}\delta_{ij}, \nonumber \\ 
\int\limits_0^{2\pi}\frac{d\theta}{2\pi}n_in_jn_l n_m&=\frac{1}{8}\left(\delta_{ij}\delta_{lm}+\delta_{il}\delta_{jm}+\delta_{im}\delta_{jl}\right).
\end{align}

Let us consider the contribution to the second harmonic from the first expression \eqref{ExpandFraction}. 
Specifically, at zero temperature, after the angular average, we need to consider the following integral:
\begin{align}\label{Intj2a}
&A\!=\!\frac{-\omega^2 z_\omega}{2 m^2 z_{2 \omega}}\!\! \sum_{s= \pm1}\!\int\limits_0^{\infty}\!\! \frac{k^3 d k}{\left(z_\omega^2+4 \alpha_{\mathrm{so}}^2 k^2\right)^2}
\left[2 \vartheta\left(\varepsilon_{\text{F}}-\epsilon_k-s \Delta_k\right)\right. \nonumber \\ 
&\left.-\vartheta\left(\varepsilon_{\text{F}}-\epsilon_k-s \Delta_k-\omega\right)-\vartheta\left(\varepsilon_{\text{F}}-\epsilon_k-s \Delta_k+\omega\right)\right]
\end{align}
Here, we replace $\Delta_k=\alpha_{\text{so}}k$ in the arguments of the Heaviside step-functions with $\Delta=\alpha_{\mathrm{so}} k_{\mathrm{F}}$, incorporating it into the definition of the Fermi energy, i.e. $\varepsilon_{\text{F}}-s\alpha_{\mathrm{so}} k_{\text{F}}\to\varepsilon_{\text{F}}$. The correction term resulting from this approximation is of the order $\left(\alpha_{\mathrm{so}}/v_{\text{F}}\right)^2\ll1$. Thus $A(\omega)$ becomes
\begin{align}
&A=\frac{-\omega^2 z_\omega}{z_{2 \omega}}\nonumber \\  
&\times\sum_{s= \pm1}\left(\int\limits_{\varepsilon_{\text{F}}-\omega}^{\varepsilon_{\text{F}}} \frac{\xi d \xi}{\left(8 m \alpha_{\mathrm{so}}^2 \xi+z_\omega^2\right)^2}-\int\limits_{\varepsilon_{\text{F}}}^{\varepsilon_{\text{F}}+\omega} \frac{\xi d \xi}{\left(8 m \alpha_{\mathrm{so}}^2 \xi+z_\omega^2\right)^2}\right),
\end{align}
where the indefinite integral has the following tabulated expression
\begin{align}
&\int \frac{\xi d \xi}{\left(8 m \alpha_{\mathrm{so}}^2 \xi+z_\omega^2\right)^2}=\nonumber \\ 
&\frac{1}{\left(8 m \alpha_{\mathrm{so}}^2\right)^2}\left[\ln \left(8 m \alpha_{\mathrm{so}}^2 \xi+z_\omega^2\right)+\frac{z_\omega^2}{8 m \alpha_{\mathrm{so}}^2 \xi+z_\omega^2}\right].
\end{align}
Using this expression and expanding in the small parameter $\omega / \varepsilon_{\text{F}} \ll 1$ we get to the leading order 
\begin{equation}
A(\omega)\approx\frac{2\omega^4 z_\omega \left(z_\omega^2-4 \Delta^2\right)}{z_{2 \omega}\left(z_\omega^2+4 \Delta^2\right)^3} + \mathcal{O}\left(\frac{\omega^4}{\varepsilon_{\text{F}}^4}\right).
\end{equation}

At the same time, the contribution from the second term in \eqref{ExpandFraction} is significantly smaller than that of the first due to the summation over the band index. Specifically, if we were to make the same approximation as before, $\veps_{\text{F}}-s\alpha_{\textrm{so}}k\to\veps_{\text{F}}$, the summation over $s$ would cancel out the corresponding contributions. Thus, instead of using this approximation, an additional expansion in powers of $\alpha_{\text{so}}/v_{\text{F}}$ is required, rendering the entire expression $(\alpha_{\text{so}}/v_{\text{F}})^2$ smaller than the contribution from the first term in \eqref{ExpandFraction}. After gathering all relevant terms, we obtain formula \eqref{j2a} in the main text.

\section{Details of the calculation for ${\mathbf j}_{2\omega}^{(\textrm{b})}$}\label{app:B}

We start with the expression for $\hat{W}$ defined in Eq. (\ref{Trw2b} and expand the ratio 
\beg\label{ratio}
\frac{{\mathbf b}_\bk}{b_\bk}\approx\frac{\alpha_{\textrm{so}}[\bk\times\hat{\mathbf z}]+g\mu_{\text{B}}{\mathbf H}}{\alpha_{\textrm{so}}k
+g\mu_{\text{B}}({\mathbf H}\cdot[\bn\times\hat{\mathbf z}])}
\en
up to the linear order in ${\mathbf H}$. In addition, we have to expand the arguments of the distribution functions. 
We therefore have to consider two different contributions to the current density. The first one can be written as follows:
\begin{align}\label{Jb1}
&{\mathbf J}=-\frac{2g\mu_{\text{B}}e^2}{m\omega^2z_\omega z_{2\omega}} 
\nonumber \\ 
&\times\int\limits_0^{2\pi}\frac{ d\theta}{2\pi}\bn\left[(\bn\cdot{\mathbf E})\left([{\mathbf E}\times\hat{\mathbf z}]\cdot{\mathbf H}\right)-(\bn\cdot{\mathbf E})^2\left([\hat{\mathbf z}\times {\mathbf H}]\cdot\bn\right)\right]
\nonumber \\ 
&\times \sum\limits_{s=\pm}\frac{s}{m}\int\limits_0^\infty\frac{k^2dk}{2\pi}\left[2\vartheta(\veps_{\text{F}}-\eps_k-s\alpha_{\textrm{so}}k)\right. 
\nonumber \\ 
&-\left.\vartheta\left(\veps_{\text{F}}-\eps_k-s\alpha_{\textrm{so}}k-{\omega}\right)-\vartheta\left(\veps_{\text{F}}-\eps_k-s\alpha_{\textrm{so}}k+{\omega}\right)\right].
\end{align}
As a next step we integrate over $k$ and sum over $s$, which brings an overall factor $(8m/\pi)(\alpha_{\text{so}}\omega^2/\eps_{\text{F}})$. Since the value of this integral includes the Fermi energy in the denominator, the contribution of ${\mathbf J}_{\textrm{b}1}$ to the second harmonic of the current density can be neglected, as it becomes of the order $O(g\mu_{\text{B}}H/\veps_{\text{F}})$.

The remaining contribution to $\textrm{Tr}_\sigma\left[\hat{W}_{\bk\eps}^{(2,\textrm{b})}\right]$ in Eq. (\ref{Trw2b}), originates from expanding the arguments of the distribution function up to the linear order in ${\mathbf H}$. As a result of this, in the limit of $T\to 0$ the integration over $k$ significantly simplifies. For example, we have
\begin{align}\label{IntkJ2b}
&\sum\limits_{s=\pm}{s}\int\limits_0^\infty\frac{k^4dk}{2\pi}\left[2\delta(\veps_F-\eps_k-s\alpha_{\textrm{so}}k)\right.
\nonumber \\ 
&\left.-\delta\left(\veps_{\text{F}}-\eps_k-s\alpha_{\textrm{so}}k-{\omega}\right)-\delta\left(\veps_{\text{F}}-\eps_k-s\alpha_{\textrm{so}}k+{\omega}\right)\right]
\nonumber \\ 
&\propto\alpha_{\textrm{so}}\frac{\omega^2}{\veps_{\text{F}}}.
\end{align}
Thus we conclude that this contribution will also be of the order of $O(g\mu_{\text{B}}H/\veps_{\text{F}})$ and therefore can be neglected. 

\section{Details of the calculation for ${\mathbf j}_{2\omega}^{(\textrm{c})}$}\label{app:C}

Following the discussion below Eq. \eqref{Trw2c} in the main text, we consider the following integral 
\begin{align}
&C\!=\!\frac{-\omega^2}{2 m^2 z_\omega z_{2 \omega}}\! \sum_{s= \pm1}\!\!\int\limits_0^{\infty}\!\! \frac{ \alpha_{\mathrm{so}}^2 k^5 d k}{\left(z_\omega^2+4 \alpha_{\mathrm{so}}^2 k^2\right)^2}\left[2 \vartheta\left(\varepsilon_{\text{F}}-\epsilon_k-s \Delta_k\right)\right. \nonumber \\ 
&\left.-\vartheta\left(\varepsilon_{\text{F}}-\epsilon_k-s \Delta_k-\omega\right)-\vartheta\left(\varepsilon_{\text{F}}-\epsilon_k-s \Delta_k+\omega\right)\right]
\end{align}
Here, we again replace $\Delta_k=\alpha_{\text{so}}k$ in the arguments of the distribution functions with $\Delta=\alpha_{\mathrm{so}} k_{\mathrm{F}}$, incorporating it into the definition of the Fermi energy. Thus $C(\omega)$ becomes
\begin{align}
&C=\frac{-\omega^2 2m \alpha_{\mathrm{so}}^2}{z_\omega z_{2 \omega}}\nonumber \\ 
&\sum_{s= \pm1}\left(\int\limits_{\varepsilon_{\text{F}}-\omega}^{\varepsilon_{\text{F}}} \frac{\xi^2 d \xi}{\left(8 m \alpha_{\mathrm{so}}^2 \xi+z_\omega^2\right)^2}-\int\limits_{\varepsilon_{\text{F}}}^{\varepsilon_{\text{F}}+\omega} \frac{\xi^2 d \xi}{\left(8 m \alpha_{\mathrm{so}}^2 \xi+z_\omega^2\right)^2}\right). 
\end{align}
As a next step, we use the tabulated integral
\begin{align}
&\int \frac{\xi^2 d \xi}{\left(8 m \alpha^2 \xi+z_\omega^2\right)^2}=-\frac{1}{\left(8 m \alpha^2\right)^3}\nonumber \\ 
&\times\left[2z_\omega^2\ln\left(8 m \alpha^2 \xi+z_\omega^2\right)+\frac{z_\omega^4}{8 m \alpha_{so}^2 \xi+z_\omega^2}-8 m \alpha^2 \xi\right],
\end{align} 
and expand the result in the small parameter $\omega / \varepsilon_{\text{F}} \ll 1$ to get
\begin{equation}
C(\omega)\approx\frac{4\omega^4 z_\omega\Delta^2}{z_{2 \omega}\left(z_\omega^2+4 \Delta^2\right)^3}.
\end{equation}

In complete analogy, we consider another term  
\begin{align}
&C^{\prime}=\frac{-\omega^2}{2 m^2 z_\omega z_{2 \omega}} \sum_{s= \pm1}\int\limits_0^{\infty} \frac{k^3 d k}{z_\omega^2+4 \alpha_{\mathrm{so}}^2 k^2}\left[2 \vartheta\left(\varepsilon_{\text{F}}-\epsilon_k-s \Delta_k\right)\right.\nonumber \\ 
&\left.-\vartheta\left(\varepsilon_{\text{F}}-\epsilon_k-s \Delta_k-\omega\right)-\vartheta\left(\varepsilon_{\text{F}}-\epsilon_k-s \Delta_k+\omega\right)\right],
\end{align}
which can be reduced to 
\begin{align}
&C^{\prime}=\frac{-\omega^2 }{z_\omega z_{2 \omega}} \nonumber\\ 
&\sum_{s= \pm1}\left(\int\limits_{\varepsilon_{\text{F}}-\omega}^{\varepsilon_{\text{F}}} \frac{\xi d \xi}{8 m \alpha_{\mathrm{so}}^2 \xi+z_\omega^2}-\int\limits_{\varepsilon_{\text{F}}}^{\varepsilon_{\text{F}}+\omega} \frac{\xi d \xi}{8 m \alpha_{\mathrm{so}}^2 \xi+z_\omega^2}\right). 
\end{align}
The indefinite integral has the following expression
\begin{align}
&\int \frac{\xi d \xi}{8 m \alpha_{\mathrm{so}}^2 \xi+z_\omega^2}=\nonumber \\ 
&-\frac{1}{\left(8 m \alpha^2\right)^2}\left[z_\omega^2\ln \left(8 m \alpha^2 \xi+z_\omega^2\right)-8 m \alpha^2 \xi\right].
\end{align}
From here we obtain after expanding in $\omega / \varepsilon_{\text{F}} \ll 1$ to get
\begin{equation}
C^{\prime}(\omega)\approx\frac{2\omega^4 z_\omega}{z_{2 \omega}\left(z_\omega^2+4 \Delta^2\right)^2}.
\end{equation}

Collecting all these contributions together we find the expression Eq. \eqref{j2wtotalFIN} in the main text.

\bibliography{biblio}

\end{document}